\documentclass{elsart}
\usepackage{graphics,graphicx,epsfig}
\usepackage{amssymb}
\begin{document}
\begin{frontmatter}
\newcommand{\ri}{ | 1\rangle}
\newcommand{\ro}{| 0\rangle}
\newcommand{\li}{\langle 1 |}
\newcommand{\lo}{\langle 0 |}
\newcommand{\lp}{\langle + |}
\newcommand{\rp}{| +\rangle}
\newcommand{\be}{\begin{equation}}
\newcommand{\ee}{\end{equation}}
\newcommand{\ra}{ | \alpha\rangle}
\newcommand{\la}{\langle \alpha |}
\newcommand{\rb}{ | \beta\rangle}
\newcommand{\lb}{\langle \beta |}
\title{Testing quantum correlations with nuclear probes}
\author{S. Hamieh$^a$, H.J. W\"ortche$^a$, C. B\"aumer$^b$, A.M. van den Berg$^a$}
\author {D. Frekers$^b$, M.N. Harakeh$^a$, 
J. Heyse$^{c,\,1}$, M. Hunyadi$^{a,\, 2}$}
\author {M.A. de Huu$^a$ , C. Polachic$^d$, S. Rakers$^b$, C. Rangacharyulu$^d$}
\address{$^a$Kernfysisch Versneller Instituut,
NL-9747 AA Groningen, The Netherlands}
\address{$^b$Institut f\"ur Kernphysik,
Westf\"alische Wilhelms-Universit\"at, M\"unster, Germany}
\address{$^c$Vakgroep Subatomaire en Stralingsfysica, Universiteit Gent,
B-9000 Gent, Belgium}
\address{$^d$Department of Physics and Engineering Physics,
University of Saskatchewan, Saskatoon, Sakatchewan, Canada S7N 5E2}

%
\thanks[jan_present]{Present address: IRMM, Joint Research Centre, Geel, Belgium}
\thanks[matyas_present]{Present address: Institute of Nuclear Research of the 
                                         Hungarian Academy of Sciences, H-4001 Debrecen, Hungary}
{E-mail: wortche@kvi.nl}
\begin{abstract}
We investigated the feasibility of quantum-correlation measurements in nuclear physics experiments.
In a first approach, we measured spin correlations of  singlet-spin ($^1S_0$) proton pairs, which were
generated  in  $^ 1$H(d,$^2$He) and $^{12}$C(d,$^2$He) nuclear charge-exchange reactions. The experiment 
was optimized for a clean preparation of the $^2$He singlet state and offered a $2\pi$ detection geometry 
for both protons in the exit channel. Our results confirm the effectiveness of the setup for theses studies, 
despite limitations of a small data sample recorded during the feasibility studies.  
\end{abstract}
\end{frontmatter}

\section{Introduction}
\label{Introduction}
%
Entanglement is believed to be a genuine  resource for quantum computers and quantum communication technology.  Entanglement shows up in composite quantum systems where the subsystems
do not have {\it pure} states of their own. This is a strict quantum phenomenon with no 
classical analogue. Entangled states of joint systems are {\it non-local}, meaning that the outcome of
measurements performed separately on each subsystem at {\it space-like} separation cannot be
reproduced by local-hidden-variables (LHV) models. Such non-locality can be revealed by a violation of an
inequality which any LHV model must satisfy. Such inequality is the Bell-type inequality \cite{bell}.
Experimental tests of the Bell-type inequality have so far been limited
to measurements with photons \cite{aspect_review} rather than measurements with massive Fermions   
with only one exception: a proton-spin correlation measurement performed by Lamehi-Rachti and Mittig 
(LRM) about 30 years ago \cite{lrm}. Note, that quantum non-contextuality has recently been tested with massive
Fermions in single-neutron interferometry experiment by Hasegawa {\it et al.} \cite{hasegawa_03}. However, it is well 
known that, if a theory is contextual, it is not necessarily non-local.
\par
The advantage of using massive Fermions to test Bell-type inequalities is that
the particles are well localized and the singlet state of the pair can be well defined by
measuring the internal energy of the two-proton system.
In this paper, we studied the feasibility of examining
spin-correlation measurements of proton pairs in a $^1S_0$ intermediate state generated in
$^1$H($d,^2$He)n and $^{12}$C($d,^2$He)$^{12}$B nuclear charge-exchange reactions.
By selecting events on basis of the structure of the excited state in the remaining nucleus and 
on basis of  the internal energy of the $^2$He system, we achieve a clean preparation of proton
pairs under controlled conditions. Our analysis of the experimental results described below
is compatible with the pioneering LRM experiment. However, our experimental setup, the experimental
procedure and the data analysis improved significantly in comparison to the LRM experiment with respect to
the following issues: 
\begin{itemize}
\item [a)] control of higher order multipole contamination of the singlet state,
\item [b)] control of the contamination due to randomly correlated pairs,  
\item [c)] causal separation of the proton pairs,
\item [d)] no preferred  quantization axis because of $2\pi$ detection geometry and
\item [e)] record of complete event topology.   
\end{itemize}
We will structure this paper as follows.
In the following section we give a brief overview of the  experimental arrangement and the data analysis,
 whereas special requirements and the feasibility of spin-correlation measurements are emphasized.
For details of the experimental setup and  the $^2$He analysis we refer to Ref. \cite{rakers_2002}. 
In section 3 we discuss details of the spin-correlation analysis and in section \ref{results} our results are presented.  
\section{Experimental setup}
\label{setup}
\begin{figure}[h]
\centering
\includegraphics[width=.8\textwidth] {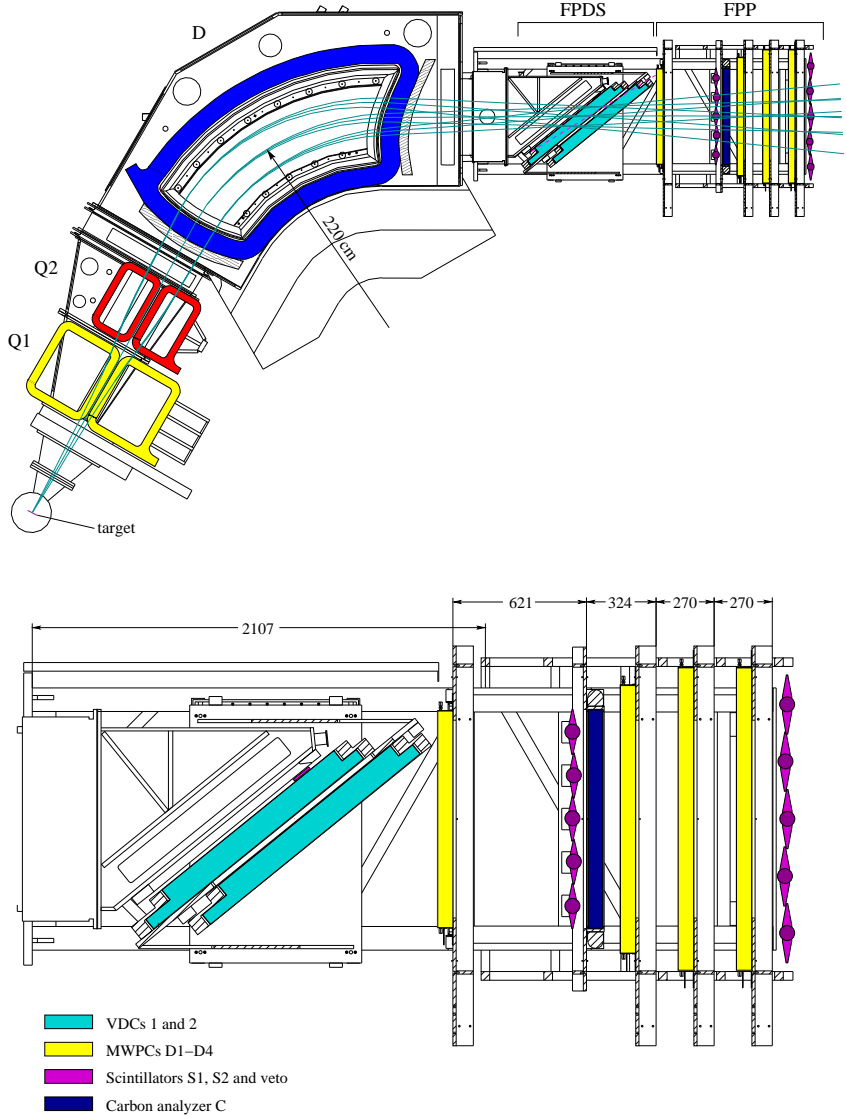}
\caption{Schematic diagram of the experimental arrangement. Top part: The BBS spectrometer in conjunction with the
ESN detector system. Lower part: Blown up view of the detector setup. The numbers indicate dimensions in units of mm.
The focal-plane detection system (FPDS) is equipped with two VDCs to determine the four-momentum vectors of protons passing the spectrometer. Further downstream, the focal-plane polarimeter (FPP)  set up, which consists of four MWPCs (D1, D2, D3, D4) and two plastic-scintillator arrrays (S1 and S2). A carbon analyzer (C) is placed just next to the scintillator S1.}
\label{setup_fig}
\vspace*{5mm}
\end{figure}
\par
The measurements were carried out using 172 MeV deuteron beams provided by the AGOR cyclotron of the Kernfysisch Versneller Instituut (KVI), Groningen. The deuterons were incident on a carbon foil of thickness 9.4 mg/cm$^2$, which was mounted in the scattering chamber of the Big-Bite Spectrometer (BBS) \cite{berg_1995}. A schematic view of the experimental setup is shown in Fig. \ref{setup_fig}. The EuroSuperNova (ESN) detector system used consists of a pair of gas-filled Vertical-Drift Chambers (VDCs) for momentum reconstruction of the protons and a focal-plane polarimeter, which comprises  Multi-Wire Proportional Chambers (MWPCs)  and a pair of scintillator paddles S1 and S2 for time-of-flight and energy-loss measurements \cite{rakers_2002}.  
\par
The $^1S_0$ proton pairs were prepared in a $^{12}$C(d,$^2$He)$^{12}$B nuclear charge-exchange reaction. A reaction of type (d,2p) is referred to as (d,$^2$He) if the outgoing protons couple to the $^1S_0$ state. The $^2$He system is unbound by an internal energy of about 0.5 MeV, which is defined by the maximum of the (pp) $^1S_0$ final-state-interaction strength. The proton pairs emerging from the charge-exchange reaction were momentum analyzed in the BBS spectrometer, which was positioned at an angle $\theta_{BBS} = 0^\circ$. The extreme forward angle was chosen to minimize the angular  momentum transfer in the reaction which favors  pure spin-flip (Gamow-Teller) type transitions and puts an additional constraint on the $^1S_0$ character of the proton pairs.
\subsection{$^2$He identification}
\begin{figure}[h]
\begin{center}
\includegraphics[width=.5\textwidth] {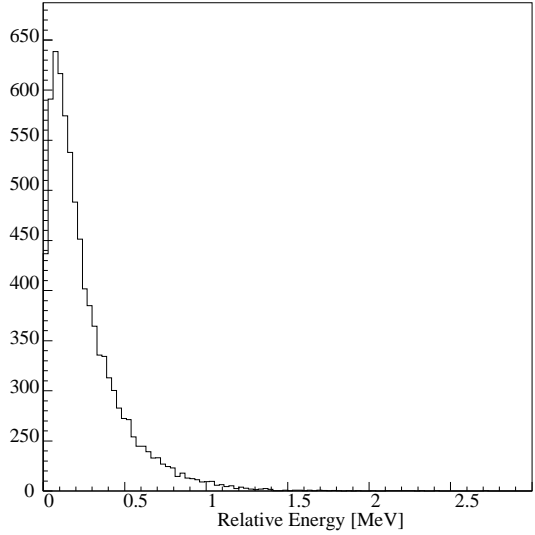}
\end{center}
\caption{Typical relative kinetic-energy distribution in the $^2$He center of mass obtained from the measured momenta
of coincident protons.}
\label{epsilon} 
\end{figure}

Due to the finite momentum and angular acceptances of the BBS focal plane, only proton pairs of relative kinetic
energies less than 1 MeV  in the $^2$He  center of mass  were detected as depicted in Fig. \ref{epsilon}. The spectrometer therefore acted as a highly exclusive filter and contaminations of higher-order multipoles were limited to the percent level \cite{kox_1993}. A dominant background of randomly correlated protons was due to the breakup of the
deuteron, a  reaction yielding protons with a momentum overlapping with the momentum range of interest.  
A clean identification of $^2$He events therefore necessitated a proper reconstruction of the excitation energy
of the residual nucleus $^{12}$B and determination of the relative timing of the two correlated protons with a good resolution.
\par
An event-trigger condition required that at least one proton passed through the spectrometer and was registered by coincident signals in scintillator planes S1 and S2.  The coincidence window was set to less than 20 ns in order to minimize  randoms caused by particles originating from different beam bursts each  separated by 23 ns.
The data, read out from the MWPCs, were fed into a fast online processing system, where they were tested on double-track conditions. Those events passing the test were stored for offline analysis.  
\par
\begin{figure}[hhh]
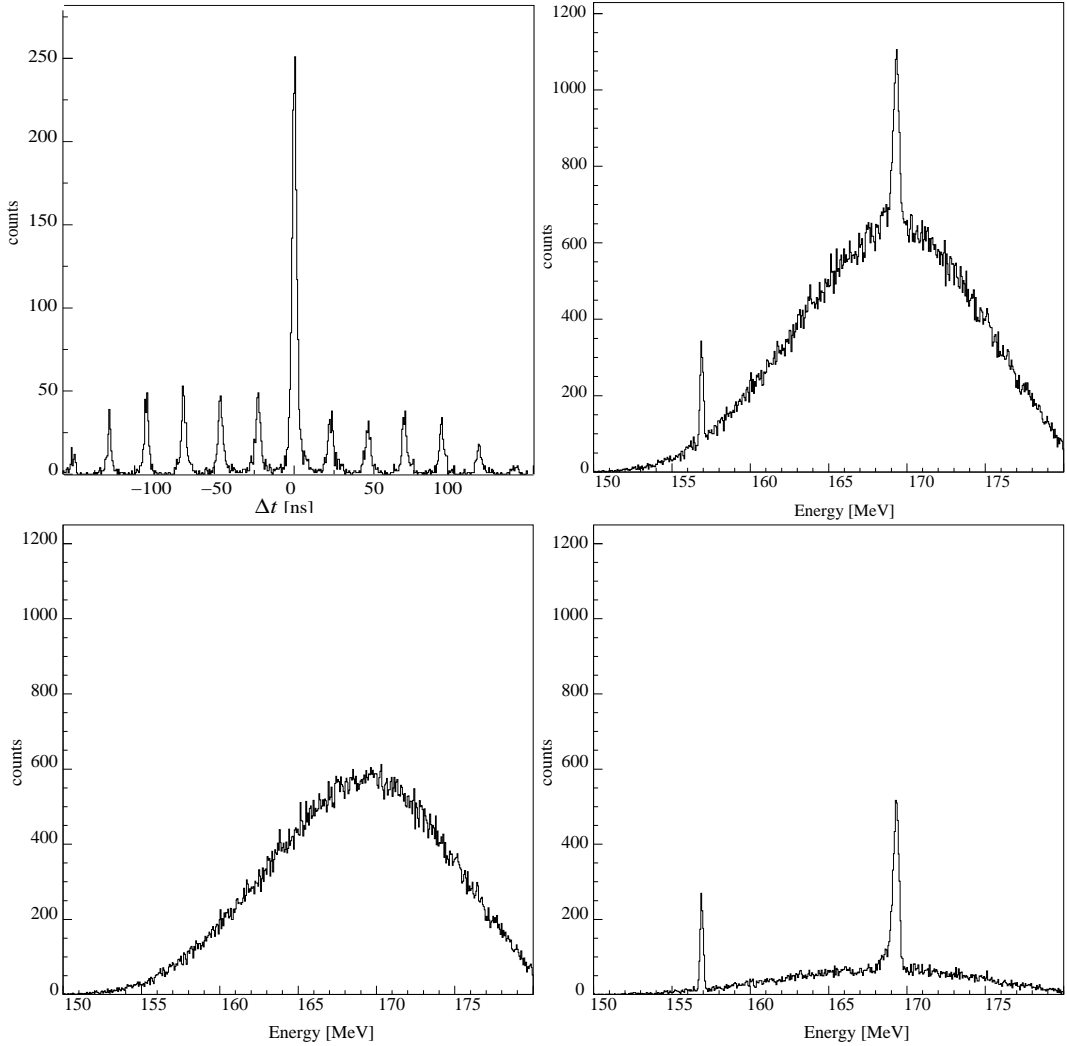

\includegraphics[width=.5\textwidth] {paw4.epsi}
\includegraphics[width=.5\textwidth] {paw1.epsi}
\includegraphics[width=.5\textwidth] {paw2.epsi}
\includegraphics[width=.5\textwidth] {paw3.epsi}
\caption{Top left: Focal-plane interception-time difference for detected proton pairs.
              For the data set shown the random contamination is about 15\%.
              Top right: Raw kinetic energy spectrum of detected proton pairs. Prominent peaks 
               indicate $^2$He protons originating from the $^{12}$B ground-state transition and 
               the p-n transition.
	         Lower left: Kinetic energy spectrum gated on satellite peaks in the interception-time difference 
	                           (no correlated protons).
	         Lower right: Kinetic energy spectrum gated on the prompt peak in the interception-time difference.
}
\label{identi}
\end{figure}

The momentum vector and the focal-plane interception time of protons were determined in offline analysis
\cite{rakers_2002} from the VDC data.
After being triggered, the VDC TDC channels remained active for about 380 ns, which is the maximum drift-time 
associated with the events.
 A typical spectrum of the difference in focal-plane interception time corresponding to kinetic energies below the $^{12}$B $Q$-value is shown in the top left of Fig. \ref{identi}. The $^2$He protons, which can for kinematical reasons be separated by at most a few $10^{-9}$ s (ns) in the BBS focal-plane, appear as a dominant {\sl prompt} peak.  The  interception-time spectrum shows a peak centered at t=0 and  satellite peaks due to random coincidences. The satellite peaks are due to randomly correlated proton pairs reflecting the beam-burst repetition rate.
An effective identification of $^2$He events could be achieved by requiring an interception-time difference in the window $\pm$10 ns. An energy spectrum accumulated under this condition is shown in the lower right of 
Fig. \ref{identi}. The kinetic energies  were calculated for two-proton events assuming $^2$He kinematics \cite{rakers_2002}.
 The energy spectrum shows two prominent peaks superimposed on a continuous  background (see top right of Fig. 
\ref{identi}). The peak at 169 MeV corresponds to the $Q$-value of the $^1$H(d,$^2$He)n reaction and is due to a 
hydrogen contamination of the $^{12}$C target. The peak at 157 MeV corresponds to the $Q$-value ($Q=-14.81$ MeV) of the $^{12}$C(d,$^2$He)$^{12}$B reaction. The structures at lower kinetic energies correspond to transitions to $^{12}$B excited states. Unphysical energies larger than in the incident beam energy of 172 MeV are due to randomly correlated protons originating from the deuteron breakup.
For the spin-correlation analysis, it was further required, that the sum of the kinetic energies of the proton pairs was equal to or less than the $^{12}$B threshold or in an energy window defined by the position and the width of the neutron peak.
%
%
\section{Spin-correlation measurements and analysis}

In order to measure the scattering angle in the carbon analyzer, identification of proton tracks upstream and
downstream of the analyzer was required. We followed the fate of each proton \cite{heyse_2002} as it passed through the carbon analyzer, acquiring the information in the detector systems
D1-D4 and S1, S2.  We  used only those  events where
both protons  scattered into an angular range larger than 3 degrees  in the carbon
analyzer, since the most forward scattering is predominantly Coulomb type which  is not
spin dependent. The experimental setup provided the flexibility to arbitrarily choose the reference axis  
during the off-line analysis.
\begin{figure}[t]
\centering
\includegraphics[width=.9\textwidth] {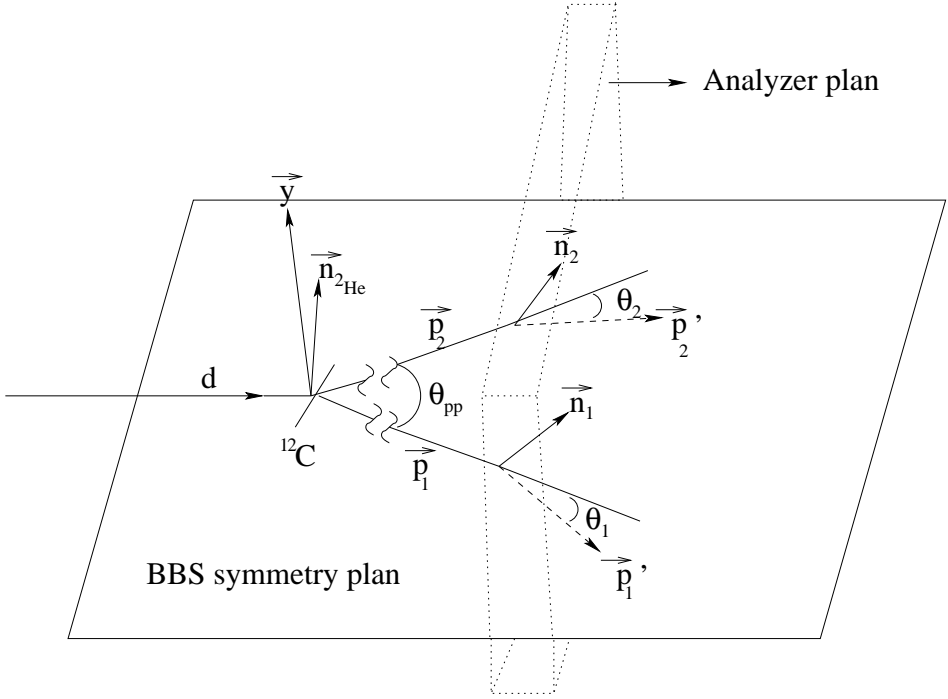}
\caption{Geometry applied for extracting the experimental correlation function Eq. \protect\ref{Pexp}.
The primary  target is indicated by $^{12}$C, the primary scattering normal $\vec{n}_{\rm ^2He}$ 
refers to the incoming deuteron and $^2$He center-of-mass movement, $\vec{y}$ is the normal to the BBS 
(horizontal) symmetry plane. The momenta of proton 1 and proton 2 upstream the analyzer are indicated 
by $\vec{p}_{1(2)}$, downstream the analyzer by $\vec{p}\,'_{1(2)}$, and the normal and angle of the analyzing scattering for proton 1 (2) are indicated by $\vec{n}_{1(2)}$
and $\theta_{1(2)}$, respectively. The angle between the proton momenta $\vec{p}_{1(2)}$ is given by $\theta_{pp}$.} 
\label{geometry}
\end{figure}
\par
The correlation function $P(\theta)$ for two spin-${1\over 2}$ states can be measured according to 
\begin{equation} 
P(\theta)  =  \frac{N_{++} + N_{--} - N_{+-} - N_{-+}}{N_{++} + N_{--} + N_{+-} + N_{-+}} 
                = \frac{N_{++} + N_{--} - N_{+-} - N_{-+}}{N_{total}} \, ,  
\end{equation}
where $\theta$ is the angle between two arbitrary quantization directions  
$\vec{Q}_1$ and  $\vec{Q}_2$ orthogonal to the momenta $\vec{p}_1$ and $\vec{p}_2$ of the two correlated protons, i.e. $\vec{Q}_1\perp \vec{p}_1$ and $\vec{Q}_2\perp \vec{p}_2$ and $N_{++}$  ($N_{--}$) is the number of events, 
where both protons scatter to the left (right) of the quantization direction and $N_{+-}$  $ (N_{-+})$ is the number of events,
where proton 1 scatters to the left (right) and proton 2 scatters to the right (left) and $N_{total}$ is the total number of events.
Taking into account the finite analyzing power of the carbon analyzer, the  number of events, in the above correlation function have to be weighted on an event-to-event basis by the analyzing powers $A^{1,2}_y(\theta_{1,2}, E_{1,2})$ \cite{McN}
\begin{equation}
N^w_{++ (--,+-,-+)} = \sum_{++(...)} \frac{1}{A^1_y(\theta_1, E_1)} \cdot \frac{1}{A^2_y(\theta_2, E_2)}
\end{equation}
yielding the experimental correlation function $P_{exp}$ (see also LRM \cite{lrm})
\begin{equation}
P_{exp} (\theta) = \frac{N_{++}^w + N_{--}^w - N_{+-}^w - N_{-+}^w}{N_{total}}\,.
\label{Pexp}
\end{equation}
\par
\begin{figure}[h]
\centering
\includegraphics[width=.7\textwidth] {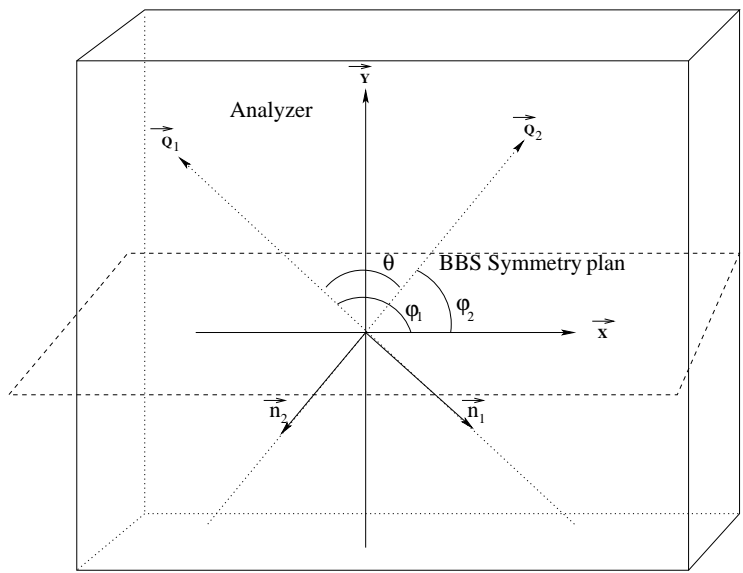}
\caption{ Sign convention for the correlations depicted for a right-right $(-\,-)$ event and $\theta_{pp}=0^{\circ}$.
 \protect\label{sign}
}
\end{figure}

The geometry applied to extract the experimental correlation function Eq. \ref{Pexp} is shown in  
Fig. \ref{geometry} and Fig. \ref{sign}. The correlation analysis was only applied for events where the direction of the 
primary scattering normal $\vec{n}_{i}$ deviated less than $1^{\circ}$ from 
the normal of the BBS symmetry plane $\vec{y}$.  Due to this selection,  our analysis 
became compatible with the LRM analysis \cite{lrm}.
\par
The sign convention for the correlations is shown in Fig. \ref{sign}, where the example of a right-right $(-\,-)$ correlation is depicted. The quantization axis $\vec{Q}_{1(2)}$ is chosen along the scattering normal  $\vec{n}_{1(2)}$ with the convention, that the projection on the normal the BBS symmetry plane $\vec{Q}_{1(2)} \cdot \vec{y}$ is positive. According to this convention, a positive definite projection of the scattering normal $\vec{n}_{1(2)}\cdot \vec{y}$ indicates scattering to the left,
a negative definite projection scattering to the right of the quantization axis.  
\par
The correlation angle $\theta$ (see Fig. \ref{sign}) is defined as
\begin{equation}
\theta = |\varphi_1 - \varphi_2|\, ,
\end{equation}
whereas the angle $\varphi_{1(2)}$ is measured in respect to the vector $\vec{x}_{1(2)}$ given by
\begin{equation} 
\vec{x}_{1,2}=\vec{y} \times \vec{p}_{1(2)} \,.
\end{equation}
The definition of  $\varphi_{1(2)}$ and $\theta$  holds also for finite $\theta_i$ because of the purely transverse character
of the analyzing reaction.
\par
In quantum theory, the operator that corresponds to the correlation function is
\begin{equation}  P=\vec{Q}_1\cdot\vec{\sigma}\otimes  \vec{Q}_2 \cdot \vec{\sigma} \,,\end{equation}
 acting in the Hilbert space ${{H_{\rm 1}}\otimes {H_{\rm 2}}}$
in $2\otimes 2$ dimension and  $\vec{\sigma}$ are the Pauli matrices.
 The correlation function $P_{QM}$ is given by the  mean value of
 this operator.
For  a pure state this correlation function could be easily computed. For a singlet state we have
 \begin{equation}
P_{QM}^{\rm Singlet}(\theta) = -\cos\theta \label{qmexpect}. \end{equation}
However, if the state is mixed the mean value
should be averaged over the ensemble. Taking into account the effect of a random contamination
of the pure singlet state,  the quantum expectation 
deviates from Eq. \ref{qmexpect}. 
In fact, we introduce a factor $\gamma$ which  interpolates between the unpolarized state
 $I/4$ and the singlet state $|\Psi^{-}\rangle=(|+-\rangle-|-+\rangle)/\sqrt(2)$, with $I$ the unit matrix.
\begin{figure}[h]
\centering
\includegraphics[width=1.\textwidth] {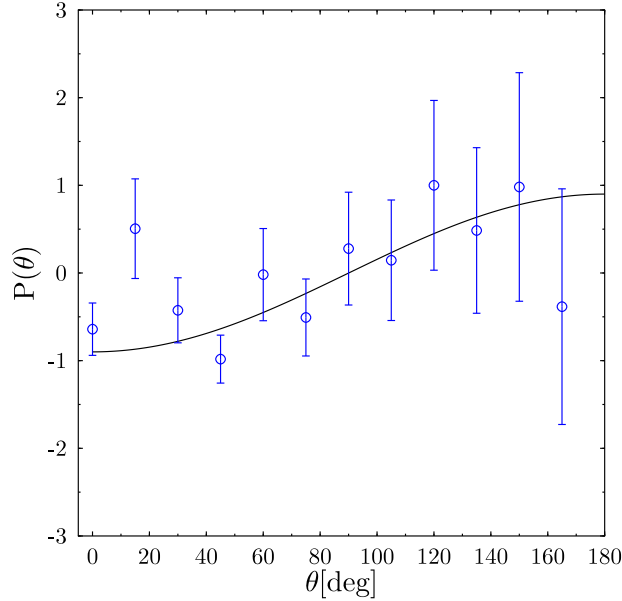}
\vspace*{-6cm}
\caption{ The extracted experimental correlation function $P_{exp}(\theta)$ (Eq. \ref{Pexp}) in comparison with  the quantum mechanics prediction $P_{QM}^{Werner}$ (Eq. \ref{werner}) for $\gamma = 0.9$}.
\label{p}
\end{figure}

\begin{figure}[h]
\centering
\includegraphics[width=.9\textwidth] {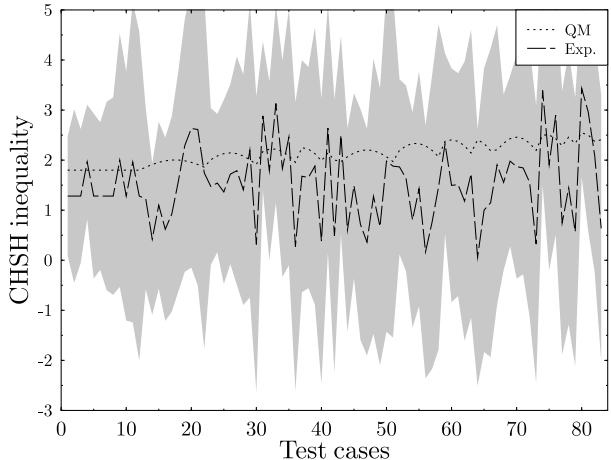}
\vspace*{-4cm}
\caption{ CHSH inequality for all possible combinations
found from the probability shown in Fig. \ref{p}. The dashed line shows the experimental results, the dotted line
the QM prediction. The shaded band represents the statistical error.
 \protect\label{chsh}
}
\end{figure}
The density matrix of such a state is called
Werner states \cite{Wern89} and it is given by
 \begin{equation} \rho_{W}=(1-\gamma){I\over 4}+\gamma|\Psi^{-}\rangle\langle\Psi^{-}|\,\end{equation}
This effect reduces the quantum expectation value of the correlation functions as  follows
 \begin{equation}
P_{QM}^{Werner}(\theta) = -\gamma\cos\theta \,. 
\label{werner}
\end{equation}
\section{Results}
During two days of data taking, the mean rate of $^2$He events ending up in the correlation analysis amounted to about 0.1 Hz/nA, despite the fact that the (d,$^2$He) production rate was nearly a factor 50 higher. 
This loss in statistics was mainly due to inefficiencies in the particle tracking downstream the analyzer. 
For the final analysis the data were binned into 12 angular $\theta$ bins, which yielded on average  $10^3$  events per angle bin.
\par
The random events, i.e. events shown in the lower left of Fig. \ref{identi}, yielded an 
averaged correlation $ \bar{P}_{exp} = 0.05\pm 0.7$ per angle, which justifies the treatment of the background as a random contribution as indicated in  Eq. \ref{werner}. The overall mean of random events contributing to the prompt events amounted to 10\%.
\par
In  Fig. \ref{p} we show the extracted  experimental correlation function $P_{exp}$ in comparison with 
the quantum mechanics prediction for Werner states Eq. \ref{werner} for $\gamma = 0.9$. Despite of the 
large uncertainties the data exhibit a trend which agrees well with the quantum-mechanics prediction and yields
a $\chi^2/{d.o.f}=\sum_i (P_{exp}^i- P_{QM}^i)/\Delta P_{exp}^i)^2 = 0.96 $, which has to be compared with 
$\chi^2/{d.o.f} = 1.93$ if $P_{QM}$  is replaced by $P_{const} = 0$.
\par
In order to demonstrate the power of the presented experimental approach, we further used the values we 
obtained for $P_{exp}$ to extract the values for a correlation function proposed by 
Clauser, Horne, Shimony and Holt (CHSH) \cite{chsh} which can be written in the form   
\begin{equation} 
|P(\theta_a, \theta_b, \theta_c)|= |P(\theta_a)-P(\theta_a+\theta_b)|+|P(\theta_a+\theta_b+\theta_c)+
                                                     P(\theta_a+\theta_c)|\,,
\end{equation}
with $\theta_{a,b,c}$ various sets of correlation angle $\theta$. All possible 
angular combinations yield 84 test cases which have been plotted in Fig. \ref{chsh} in comparison with 
the quantum-mechanics predictions. A discussion of the CHSH-type correlations is beyond the scope of the 
present paper and will be the topic of a forthcoming publication  \cite{polachic}. From Fig. \ref {chsh} it becomes obvious that the present experimental data set suffers from large statistical uncertainties.  Nevertheless, the data exhibit  a tendency to stay below the quantum mechanical results. A fact which would  point towards classical scenarios. 

\label{results}
\section{Conclusion}
\label{conclusion}
In this paper, we presented a new experimental approach to study the feasibility of examining
spin-correlations measurements in nuclear physics. With an improved detector setup, which removes
the ambiguity in the track reconstruction, measurements  with significant precision will become
feasible. We are convinced that our experiments will have many potential applications in future quantum communication 
technology.  
\par
This work was performed as part of the research program of the {\sl Stichting voor Fundamenteel Onderzoek 
der Materie (FOM)} with financial support from the {\sl Nederlandse Organisatie voor Wetenschappelijk Onderzoek }. It was 
supported by the NSERC Canada, the European Union through the Human Capital and Mobility Program and the Fund for Scientific Research (FSR) Flanders. 

\end{document}